\documentclass[12pt]{article}
\usepackage{graphicx}
\usepackage{epstopdf}
\usepackage{amsmath}
\usepackage{amssymb}
\usepackage{amsthm}
\usepackage{ccaption}
\usepackage[usenames]{color}

\usepackage[
      colorlinks=true,
      linkcolor=black,
      citecolor=black,
      urlcolor=blue,
      filecolor=blue,
      pdfstartview=FitV,
      pdftitle={entanglement entropy},
        pdfauthor={Hubeny, Rangamani},
        pdfsubject={entanglement entropy},
        pdfkeywords={min surface, disconnected, entanglement},
        pdfpagemode=None,
        bookmarksopen=true
      ]{hyperref}

\makeatletter
\@addtoreset{equation}{section}

\makeatletter
\renewcommand\section{\@startsection {section}{1}{\z@}%
                                   {-3.5ex \@plus -1ex \@minus -.2ex}
                                   {2.3ex \@plus.2ex}%
                                   {\normalfont\large\bfseries}}
\renewcommand\subsection{\@startsection{subsection}{2}{\z@}%
                                     {-3.25ex\@plus -1ex \@minus -.2ex}%
                                     {1.5ex \@plus .2ex}%
                                     {\normalfont\bfseries}}

  \captionnamefont{\bfseries}
  \captiontitlefont{\small\sffamily}
  \captiondelim{: }
  \hangcaption

\def\baselinestretch{1.2}
\newcommand{\be}{\begin{equation}}
\newcommand{\ee}{\end{equation}}
\newcommand{\beq}{\begin{eqnarray}}
\newcommand{\eeq}{\end{eqnarray}}

\def\sec#1{\S \;\ref{#1}}
\def\fig#1{Fig.\,\ref{#1}}
\def\req#1{(\ref{#1})}

\def\[{\left [}
\def\]{\right ]}
\def\({\left (}
\def\){\right )}

\def\eg{{\it e.g.}}

\def\cf{{\it cf.}}
\def\ie{{\it i.e.}}

\def\etc{{\it etc.}}

\def\btab{\begin{table}[h] \begin{center} \begin{tabular}{l lp{3in}}}
\def\etab{\end{tabular} \end{center} \end{table}}
\def\btabm{\begin{center} \begin{tabular}}
\def\etabm{\end{tabular} \end{center}}

\def\p{\partial}

\def\tot{\longleftrightarrow}

\def\CA{{\cal A}}

\def\CN{{\cal N}}

\def\CW{{\cal W}}

\def\Rb{{\bf R}}

\def\A5S5{{\rm AdS}_5 \times \S^5}

\def\p{\partial}

\def\pt{\( {\partial \over \partial t } \)}

\def\p{\partial}

\def\Tr#1{{\rm Tr}\(#1\)}

\newcommand{\bbibitem}[1]{\bibitem{#1}\marginpar{#1}}

\def\Label#1{\label{#1}%
{ \color{blue}{\smash{\hbox to0pt{\raise2ex\hbox{\tiny[#1]}\hss}}}}}
\def\noLabels{\let\Label=\label}
\def\nobbibitem{\let\bbibitem=\bibitem}

\def\bulk{{\cal M}}
\def\bdy{\p{\cal M}}

\def\Gms{\CW}

\def\rA{\CA}

\def\bs{\, \backslash \,}

\def\s#1{s_{#1}}
\def\R#1{R_{#1}}
\def\pt{p}
\def\EE{S}
\def\Cv{{ \cal C }}

\title{{\bf Holographic entanglement entropy for disconnected regions}}

\author{Veronika E Hubeny\footnote{veronika.hubeny@durham.ac.uk} \ and Mukund Rangamani\footnote{mukund.rangamani@durham.ac.uk}\\ \\
\small \sl   Centre for Particle Theory \& Department of
Mathematical Sciences,
\\[-1.5mm]
\small \sl Science Laboratories, South Road, Durham DH1 3LE, United Kingdom \\
}

\begin{document}
\noLabels 
\nobbibitem 

\setlength{\baselineskip}{16pt}
\begin{titlepage}
\maketitle
\begin{picture}(0,0)(0,0)
\put(350, 320){DCPT-07/65}
\end{picture}
\vspace{-36pt}

\begin{abstract}
We present a simple derivation  of the entanglement entropy for a region made up of a union of disjoint intervals in 1+1 dimensional quantum field theories using holographic techniques. This generalizes the results for 1+1 dimensional conformal field theories derived previously by exploiting the uniformization map.
We further comment on the generalization of our result to higher dimensional field theories.
 \end{abstract}
\thispagestyle{empty}
\setcounter{page}{0}
\end{titlepage}

\renewcommand{\baselinestretch}{1.2}  

\section{Introduction}
 \label{intro}

Entanglement entropy is an important concept in field theory systems, with applications to condensed matter systems, quantum information, \etc. Further, given its non-extensive nature \ie, area scaling  \cite{Srednicki:1993im,Kabat:1994vj}, one is tempted,  in the context of holography, to  think of it as providing a measure for the effective degrees of freedom associated with a given region. This interpretation is natural in light of the recent proposal  by Ryu \& Takayanagi \cite{Ryu:2006bv,Ryu:2006ef} to compute entanglement entropy in quantum field theories by finding the area of an appropriate bulk surface in Planck units. Further support for this prescription arises in the  covariant generalization of their proposal \cite{Hubeny:2007xt}, where it was argued that the entanglement entropy is related to  light-sheet constructions of the covariant entropy bound \cite{Bousso:1999xy} in the bulk spacetime.

Before proceeding to the technical aspects, let us recall the definition of entanglement entropy and the holographic prescription for computing it. We consider a quantum field theory on $\bdy = \CN \times \Rb_t$ and focus on a region $\rA\subset \CN$. Given a density matrix $\rho$ (or a pure state) on $\CN$ we define the reduced density matrix $\rho_\rA = {\rm Tr}_{\CN \backslash \rA} \( \rho\)$. The entanglement entropy is then simply the von Neumann entropy associated with $\rho_\rA$: $S_\rA = - \Tr{\rho_\rA \, \log \, \rho_\rA}$. The covariant holographic prescription \cite{Hubeny:2007xt} for computing this  is to consider the problem of finding a co-dimension two extremal surface $\Gms_\rA$ in the bulk geometry  $\bulk$ (dual to the state in question), which ends on the boundary of the chosen region $\rA$ at the spacetime boundary $\bdy$. $S_\rA$ is then given by the area of $\Gms_\rA$ in Planck units \ie, $S_\rA = {{\rm Area }\(\Gms_\rA\) \over 4 G_N}$. In the situation where the bulk geometry is static (and hence the dual state is time-invariant) the problem reduces to the simpler one of finding minimal area surfaces \cite{Ryu:2006bv,Ryu:2006ef} (see \cite{Fursaev:2006ih} for a proof). Recent discussions of entanglement entropy in the holographic context include braneworld scenarios \cite{Emparan:2006ni}, closed string tachyon condensation \cite{Nishioka:2006gr}, and confinement-deconfinement transitions \cite{Nishioka:2006gr, Klebanov:2007ws,Faraggi:2007fu}.

The technology to compute entanglement entropy is best developed in the case of 1+1 dimensional CFTs \cite{Calabrese:2004eu} (\cf,  \cite{Holzhey:1994we} for earlier work and \cite{Cardy:2007mb} for generalizations to integrable non-conformal systems).  In such situations one can exploit the power of 1+1 dimensional conformal invariance to compute the entanglement entropy for a region composed of a disjoint union of intervals on a spatial slice.\footnote{The trick is to consider evaluating $\Tr{\rho_\rA^n}$ using a path integral prescription. One computes $S_\rA$ by analytic  continuation and taking the derivative as $n \to 1$. This replica trick allows one to relate the computation of the entanglement entropy to the  computation of twist operator correlation functions in CFTs \cite{Calabrese:2004eu}.} However, a key feature used in this derivation, the Riemann mapping theorem,  is not available for non-conformal theories or for higher dimensional examples.

On the other hand, the holographic recipe for computing entanglement entropy, which requires us to compute the area of an extremal co-dimension two spacelike surface (or a minimal surface for static states), is much more powerful, since we only need to solve a classical problem in a given bulk geometry. As an illustrative example, this holographic relation has been used recently \cite{Headrick:2007km} to give a simple and elegant proof of the  strong sub-additivity property. The proof of this statement in standard quantum field theory is quite involved, see \cite{Lieb:1973cp,Lieb:1973lr}.

Here we use the holographic prescription to derive the expression for entanglement entropy of disconnected regions in 1+1 dimensions for time independent states (or density matrices).  We  obtain the answer predicted by the field theory analysis of \cite{Calabrese:2004eu} for CFTs, but now only using geometrical arguments.  In particular,  the definition of minimal surface, which in turn guarantees the strong sub-additivity property, plays a crucial role.  We can easily generalise the calculation to $N$ disconnected regions in $1+1$ dimensional CFT (in a static state), again giving agreement with the field theoretic predictions. However, the derivation we present is true for any 1+1 dimensional quantum field theory, not necessarily one enjoying conformal symmetry, demonstrating the power of the geometric construction.
We also generalise our calculations to higher dimensions in some particular situations. Throughout, we assume a static configuration (so we can consider bulk minimal surfaces on a constant time slice), and comment on potential generalizations in the Discussion.

We begin by deriving a formula for entanglement entropy for disconnected regions in the case of 1+1 dimensional field theories in \sec{disconnect}. We then present a conjecture for the higher dimensional case in \sec{disconnectd} and conclude with a set of open questions in \sec{discussion}. 

\section{Disconnected regions in $1+1$ dimensional field theories}
 \label{disconnect}

Consider a $1+1$ dimensional QFT, living on the boundary of a $2+1$ dimensional asymptotically-AdS bulk spacetime. We are interested in calculating the entanglement entropy of a region composed of disconnected intervals. We will label the points bounding the intervals as $p_i$ with $i = 1, \ldots, n$ and the region of interest will be 
\begin{equation}
X = \bigcup_{m=0}^{[n/2]} \, \[p_{2m+1}, p_{2m}\] .
\Label{Xregion}
\end{equation}	
Further, we will denote the simply connected region with endpoints $\pt_i$ and $\pt_j$ by $\R{ij}$, and the corresponding entanglement entropy of this region by $\EE(\R{ij})=\s{ij}$. By the holographic prescription of \cite{Ryu:2006bv,Ryu:2006ef} 
we have
\begin{equation}
s_{ij}= {1\over 4 \, G_N^{(3)}} \, {\rm Length} \(\gamma_{ij}\)
\Label{holdefs}
\end{equation}	
where $\gamma_{ij}$ is the minimal length curve (spacelike geodesic) connecting $p_i$ and $p_j$ through the bulk.\footnote{The curves under consideration should be homologous to the boundary region of interest \ie, one must be able to deform the bulk curve back to the boundary without any obstruction.}  Note that the entanglement entropy of a region consisting of a single point vanishes, so that $\s{ii}=0$.   Furthermore, we don't require any orientation, \ie,  $\R{ij}= \R{ji}$ and correspondingly $\s{ij}=\s{ji}$.  Hence we have $n(n-1)/2$ such regions.  

The quantity of interest will be the entanglement entropy of a non-simply connected region $X$, and we will seek a universal expression for $x \equiv \EE(X)$ in terms of the entanglement entropies of related simply connected regions, namely the $\s{ij}$'s. 

\subsection{Two disconnected regions}
 \label{disconnecttwo}

\begin{figure}
\begin{center}
 \includegraphics[width=14cm]{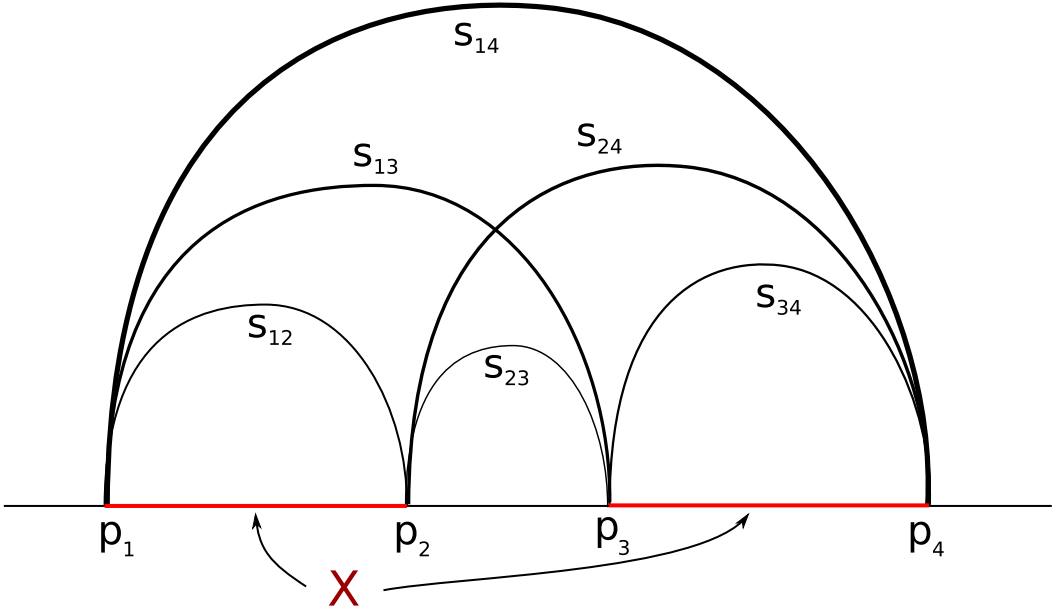}
\caption{Minimal surfaces $\gamma_{ij}$ in the bulk, whose length gives the entanglement entropy $\s{ij}$, corresponding to the boundary intervals $\R{ij}$ bounded by points $\pt_i$ and $\pt_j$. The disconnected boundary region $X$ is indicated in red.} 
\label{minsurfX}
\end{center}
\end{figure}

Let us now concentrate of the case of two disconnected regions, $n=4$; the case of general $n$ will be considered in the next subsection.
The six minimal surfaces\footnote{Given the linear relation between ${\rm Length}(\gamma_{ij})$ and $\s{ij}$, we will henceforth use $\s{ij}$  to denote the entanglement entropy as well as the bulk minimal surface.} for the $n=4$ case are sketched in \fig{minsurfX}.

We can construct the disconnected region of interest $X$ either as union or equivalently as a difference of two regions:
\begin{equation}
X = \R{12} \cup \R{34} = \R{14} \bs \R{23}
\Label{XdefR}
\end{equation}	
Our main objective is to determine $x=\EE(X)$ in terms of $\s{ij}$.   We will achieve this by constraining $x$ using the minimal area property of each $\s{ij}$ and strong sub-additivity.

\paragraph{Minimal surface and sub-additivity inequalities:} Let us start by writing the inequalities obeyed by the various minimal surfaces  $\s{ij}$.
Consider two endpoints, $\pt_i$ and $\pt_j$ and consider any bulk `surface' (in this $1+1$ dimensional case just a curve) ending on these endpoints.  By definition of minimal surface, $\s{ij}$ must be the smallest of all such surfaces.  In particular, the minimal surface $\s{ij}$ cannot have greater area than the surface composed of two minimal surfaces $\s{ik}$ and $\s{jk}$ which meet at the same point $\pt_k$ on the boundary. This implies the set of inequalities of the form\footnote{Strictly speaking, in  any local quantum field theory, the entanglement entropy is divergent due to contributions from the degrees of freedom localized near the boundaries. In the holographic computation this is manifested by the divergent area of any bulk surface in asymptotically AdS spacetimes. We will assume henceforth that we are working with a fixed regulator to make sense of the inequalities. Alternately, one can imagine working with an appropriate background subtraction scheme.}
\begin{equation}
\s{ij} \le \s{ik} + \s{jk}
\Label{sijk}
\end{equation}	
for $i,j,k = 1, \ldots, 4$. 
Since this is non-trivial only when $i,j$ and $k$ are all distinct, and since $\s{ij}$ is symmetric, we have a-priori $12$ independent inequalities.

However, this set of inequalities can be substantially shortened by using the property of strong sub-additivity \cite{Lieb:1973cp} (see \cite{Lieb:1974qr,Araki:1970ba} for excellent discussions of entropy inequalities in classical and quantum systems).
For arbitrary regions $A$ and $B$, we have the two basic strong sub-additivity inequalities:
\begin{equation}
S(A) + S(B) \ge S(A \cup B) + S(A \cap B)  
\Label{subaddgenABcup}
\end{equation}	
\begin{equation}
S(A) + S(B) \ge S(A \bs B) + S(B \bs A)
\Label{subaddgenABmin}
\end{equation}	
In \cite{Headrick:2007km} it was  shown that these inequalities also follow immediately from the definition of minimal surfaces (\cf,  \cite{Hirata:2006jx} for early discussions of sub-additivity in the holographic prescription and \cite{Fursaev:2007sg} for recent discussions).  To see this, consider the minimal surfaces $\s{ij}$ shown in \fig{minsurfX}.  There are exactly two pairs of surfaces ending on distinct endpoints which don't intersect, namely $(\s{12},\s{34})$ and $(\s{23},\s{14})$, and there is one pair of intersecting surfaces, $(\s{13},\s{24})$.  We can now decompose the latter pair in two different ways, forming two non-minimal surfaces glued together at the intersection.  Using the same argument as above, the sum of the areas of these two non-minimal surfaces cannot be smaller than the sum of the corresponding minimal surfaces.  This yields the two stronger inequalities  \cite{Headrick:2007km}:
\begin{equation}
\s{14} + \s{23} \le \s{13} + \s{24}
\Label{subaddcup}
\end{equation}	
\begin{equation}
\s{12} + \s{34} \le \s{13} + \s{24}
\Label{subaddmin}
\end{equation}	
Note that we can obtain \req{subaddcup} and \req{subaddmin} from \req{subaddgenABcup} and \req{subaddgenABmin}, respectively, by taking $A = \R{13}$ and $B= \R{24}$.

Each of \req{subaddcup} and \req{subaddmin} combines with two equations of the form \req{sijk} to yield two of the other equations of the form \req{sijk}, rendering the latter set redundant.  For example, adding 
$\s{ij} + \s{kl} \le  \s{ik} + \s{jl} $ and 
$\s{ik} \le  \s{ij} + \s{jk} $ yields 
$\s{kl} \le  \s{jl} + \s{jk} $.
Eliminating the redundant equations in this manner, we reduce the 12 original equations \req{sijk} to the following four:
\begin{eqnarray}
\s{13}  &\le& \s{12} + \s{23} \cr
\s{13}  &\le& \s{14} + \s{34} \cr
\s{24}  &\le& \s{12} + \s{14} \cr
\s{24}  &\le& \s{23} + \s{34} 
\Label{sijkmin}
\end{eqnarray}	
The set of the six inequalities \req{subaddcup}, \req{subaddmin}, and \req{sijkmin} forms the minimal set of independent constraints on the minimal surfaces $\s{ij}$.  
Indeed, we can easily check that there exist examples of $\s{ij}$ satisfying any five of the six inequalities but not the sixth one.

\paragraph{Constraints on $x$:} We now turn to constraining the entanglement entropy $x = \EE(X)$ of the disconnected region $X$.  The strategy is to use the strong sub-additivity property \req{subaddgenABcup} and \req{subaddgenABmin} for suitable regions $A$ and $B$ such that one of the terms corresponds to $\EE(X)$.  There are only two nontrivial possibilities for obtaining an upper bound on $x$, which yield:
\begin{equation}
x \le \s{12} + \s{34}
\Label{xupcup}
\end{equation}	
\begin{equation}
x \le \s{14} + \s{23}
\Label{xupmin}
\end{equation}	
These follow from
\req{subaddgenABcup} with $A=\R{12}$ and $ B=\R{34}$  and from
\req{subaddgenABmin} with $A=\R{14}$ and $ B=\R{23}$, respectively.
The possibilities for obtaining lower bounds on $x$ are more numerous, since here we can simply let $A=X$ and consider all other regions as canditates for $B$. 
A-priori, we can let $B$ be any of the six simple regions $\R{ij}$, and for each case, we have the two constraints \req{subaddgenABcup} and \req{subaddgenABmin}.  However, it turns out that four of the twelve resulting inequalities are trivial, and four more are redundant (in particular they follow from the remaining four constraints and \req{sijkmin}).
The four nontrivial independent lower bounds on $x$ which remain are the ones obtained from \req{subaddgenABcup} and \req{subaddgenABmin} with $A=X$ and $B= \R{13}, \R{24}$:
\begin{eqnarray}
\s{12} + \s{14} - \s{13}  &\le& x \cr
\s{14} + \s{34} - \s{24}  &\le& x \cr
\s{12} + \s{23} - \s{24}  &\le& x \cr
\s{23} + \s{34} - \s{13}  &\le& x 
\Label{xlowbnd}
\end{eqnarray}	
The six inequalities \req{xupcup}, \req{xupmin}, and \req{xlowbnd} form the minimal set of independent constraints on $x$.  
As we will see, how strongly these bounds constrain the actual value of $x$ in any particular case depends on how nearly the inequalities \req{subaddcup}, \req{subaddmin}, and \req{sijkmin}  are saturated.   However, as we will argue below, it is easy to construct specific examples where two of the six inequalities in \req{subaddcup}-\req{sijkmin} are saturated, thereby forcing the lower and upper bounds on $x$ from \req{xupcup}-\req{xlowbnd} to actually coincide.

We now wish to use these six constraints on $x$, along with the six further constraints on the minimal surfaces $\s{ij}$, to determine $x$ completely in terms of $\s{ij}$.  We begin by observing that there is a natural pairing between the inequalities such that each pair contains all minimal surfaces exactly once.  For example, \req{subaddcup} is paired with \req{xupcup}, \req{subaddmin} is paired with \req{xupmin}, the first of \req{sijkmin} is paired with the second of \req{xlowbnd}, \etc.  Moreover, the relations between the coefficients of $\s{ij}$ are everywhere the same.  In particular, $\s{12}$, $\s{14}$, $\s{23}$, and $\s{34}$ appear with coefficient $+1$ while $\s{13}$ and $\s{24}$ appear with coefficient $-1$.  This is strongly suggestive of a specific pattern, which we now proceed to unearth.

\paragraph{Deriving $x(\s{ij})$:} Assuming that $x$ can be expressed in terms of the minimal surfaces $\s{ij}$, we will first write a simple ansatz for the form of $x$ and then use specific limits of the various regions to fix $x$ completely.  The bounds on $x$ suggest a linear relation between $x$ and the $\s{ij}$, namely:
\begin{equation}
x = c_{12} \, \s{12} + c_{13} \, \s{13} + c_{14} \, \s{14} 
   + c_{23} \, \s{23} + c_{24} \, \s{24} + c_{34} \, \s{34} 
\Label{xlinansatz}
\end{equation}	
where the $c_{ij}$ are some constant coefficients to be determined.
In particular, according to our ansatz, $x$ is given by the expression \req{xlinansatz} for {\it any} allowed set of values of the $\s{ij}$.
Let us now consider specific limits where we know $x$ explicitly, namely the limits in which $X$ reduces to a simply connected region.  There are three nontrivial possibilities: we can take $\pt_{1} \to \pt_{2}$, $\pt_{2} \to \pt_{3}$, or $\pt_{3} \to \pt_{4}$.  In the first case, $\s{12} \to 0$ and $X \to \R{34}$, so that $x \to \s{34}$.  Furthermore, $\s{13} \to \s{23}$ and $\s{14} \to \s{24}$.  The expression \req{xlinansatz} then implies that $c_{34} = 1$ while $c_{13} + c_{23} = 0$ and $c_{14} + c_{24} = 0$.
Similarly, for $\pt_{2} \to \pt_{3}$, we have $x \to \s{14}$ 
while $\s{12} \to \s{13}$ and $\s{24} \to \s{34}$; 
hence \req{xlinansatz} yields $c_{14} = 1$, $c_{12} + c_{13} = 0$, and $c_{24} + c_{34} = 0$.  Finally, for $\pt_{3} \to \pt_{4}$, we have $x \to \s{12}$ and therefore $c_{12} = 1$, $c_{13} + c_{14} = 0$, and $c_{23} + c_{24} = 0$.  Taken together, these determine all coefficients fully; in particular we find $c_{12} = c_{14} = c_{23} = c_{34} = -c_{13} = -c_{24} = 1$.  This therefore implies that $x$ is in general given by 
\begin{equation}
x = \s{12} + \s{23} + \s{34} +  \s{14}  - \s{13} - \s{24} \ .
\Label{xanswer}
\end{equation}	
To recap, using the linear ansatz \req{xlinansatz} we have shown that the formula for the entanglement entropy of two disconnected regions is indeed given by the expression we would have naturally derived from the CFT \cite{Calabrese:2004eu}, namely \req{xanswer}.  In fact, we can  generalise \req{xlinansatz} to a sum of arbitrary functions of one variable and still derive the same result \req{xanswer}, although a completely arbitrary function of the six $\s{ij}$'s is not uniquely fixed by three special limits at hand, as can be seen by Taylor expanding and solving for the coefficients at each order.

The preceding argument establishes the desired result by making a specific (albeit sensible) ansatz for the form of the entanglement entropy for disconnected regions. A-priori one might have hoped that the inequalities derived earlier \req{xupcup}-\req{xlowbnd}, along with the constraints \req{subaddcup}-\req{sijkmin}, would suffice to derive the form of $x$. This is generically not possible without making further assumptions. 
However, if one makes the natural assumption that $x$ will be given by a unique expression (not necessarily linear) in terms of the six $\s{ij}$'s for {\it any} set of values $\s{ij}$ which satisfy the constraints \req{subaddcup}-\req{sijkmin}, then one can use a clever choice of specific `extremal' allowed values of the $\s{ij}$'s to derive the expression \req{xanswer}.
In particular, by explicitly assuming that the minimal surfaces saturate one of the inequalities \{\req{subaddcup},\req{subaddmin}\} and a corresponding one of \req{sijkmin}, we can obtain the desired expression for $x$ by showing that the upper and lower bounds on $x$ from \req{xupcup}-\req{xlowbnd} actually coincide.
For example, suppose that \req{subaddcup} and the first of \req{sijkmin} are saturated;\footnote{
Note that the system with one upper bound constraint and one lower bound constraint being saturated is perfectly self-consistent, in the sense that all of the constraints on $\s{ij}$ remain satisfied.} we can easily obtain (for arbitrary constants $a_1$ and $a_2$)
\begin{equation}
\s{14} + \s{34} - \s{24} + \( \s{12} + \s{23} - \s{13} \)  a_1 
\le x \le
\s{12}  + \s{34} + \( \s{14} + \s{23} - \s{13} - \s{24} \)  a_2
\Label{xuplobnd}
\end{equation}	
which (by setting $a_1=a_2=1$) yields \req{xanswer}. 
While this particular choice of $a_1$ and $a_2$, as well as the specific pair constraints to be saturated, may seem somewhat ad-hoc, it turns out that {\it any} choice which leads to upper and lower bounds on $x$ coinciding in fact yields \req{xanswer} uniquely.

Hence this argument may be regarded as an alternative derivation of the desired result \req{xanswer}, which uses a different set of starting assumptions.  
In particular, here we only needed to assume that $x$ is given by some universal expression in terms of the  $\s{ij}$, without assuming anything about the form of this expression.  However, in contrast to the previous argument, we required that this universal expression holds for any set of $\s{ij}$ which satisfy the constraints.\footnote{
Note that in the preceding argument, we required our linear ansatz to be satisfied by any {\it allowed} configuration of the $\s{ij}$'s; the set of such configurations is a subset of the all configurations satisfying the constraints.
Although these constraints comprised a complete set of properties derivable from the definition of a minimal surface, it is not clear whether the extremal cases used above are actually realised any physical system.}
The two arguments in conjunction therefore strengthen our derivation of \req{xanswer} by weakening the assumptions.

\subsection{Multiply disconnected regions}
 \label{disconnectmore}

Having considered the case of $X$ being composed of two disjoint intervals in the previous section, let us now turn to the more general case of non-simply connected regions, namely $m$ disjoint intervals.\footnote{ 
In one spatial dimension, the most general non-simply connected region consists of $m$ disjoint intervals for any $m \ge 2$ (in this discussion, we will ignore fractal regions such as Cantor sets, \etc); as we will see in the next section, in higher dimensions we have many more options.
}
In particular, these are bounded by the $n = 2m$ distinct endpoints $\pt_i$, which
define $n(n-1)/2$ distinct simple intervals $\R{ij}$, each associated to the entanglement entropy given by the minimal surfaces $\s{ij}$.  As a direct extension of \req{XdefR}, we have 
\begin{equation}
X = \R{12} \cup \R{34} \cup \cdots \cup \R{n-1,n} 
= \R{1n} \bs \R{23} \bs \R{45} \bs \cdots \bs \R{n-2,n-1} 
\Label{XdefRn}
\end{equation}	

Again, by definition of minimal surface, we have $\s{ij} \le \s{ik} + \s{jk}$ for $i,j,k =1, \ldots, n$; so there are a-priori $n(n-1)(n-2)/2$ independent inequalities.
Similarly, we have additional (strong sub-additivity) inequalities, which can be derived by considering any two intersecting surfaces.
For any set of four endpoints, say $\pt_i < \pt_j < \pt_k < \pt_l$, we have exactly two pairs of surfaces ending on distinct endpoints which don't intersect, namely $(\s{ij},\s{kl})$ and $(\s{jk},\s{il})$, and one pair of intersecting surfaces, $(\s{ik},\s{jl})$.  As before, we can break up the intersecting minimal surfaces into two sets of non-minimal surfaces, obtaining the inequalities of the form
\begin{eqnarray}
\s{ik}  + \s{jl} &\ge& \s{ij} + \s{kl} \cr
\s{ik}  + \s{jl} &\ge& \s{il} + \s{jk}
\Label{sijkln}
\end{eqnarray}	
For $n$ boundary points, we have ${ 1 \over 12} \, n \, (n-1) \, (n-2) \, (n-3)$ of these stong sub-additivity type inequalities.  (Note that this grows as a quartic, which is more rapid than the cubic we obtained for the simple inequalities.)  In addition, we can consider still further inequalities by generalising this argument.  For example, for six distinct endpoints, $\pt_i < \pt_j < \pt_k < \pt_l < \pt_m < \pt_n$, we can consider the pairwise intersecting surfaces, and obtain inequalities of the form\footnote{These general entropy inequalities are a natural analog of strong sub-additivity for multiple disconnected regions. As far as we are aware they have not been considered in the field theory literature before.}
\begin{equation}
\s{ik} + \s{jm} + \s{ln} \ge \s{in} + \s{jk} + \s{lm} \ , \qquad \etc.
\Label{sijklmnn}
\end{equation}	
Thus finding the minimal set of independent constraints on the minimal surfaces $\s{ij}$, not to mention on $x$, is a daunting prospect in general. 

However, we can use the same trick as in \sec{disconnectmore} to find the general form of $x$ using the previous result \req{xanswer}.
In particular, assume that $x$ is given by an expression of the form
\begin{equation}
x = \mathop{ \sum_{{i,j=1}}^n}_{j>i} \, c_{ij} \, \s{ij} 
\Label{xlinansatzn}
\end{equation}	
for some fixed constants $c_{ij}$.
We require that $x$ reduce to the correct value for all the various combinations of $\s{ij}$ vanishing.  For example, if all but 4 endpoints $\pt_i < \pt_j < \pt_k < \pt_l$ coincide so that $X = \R{ij} \cup \R{kl}$, then \req{xlinansatzn} must reduce to 
$x = \s{ij} + \s{jk} + \s{kl} + \s{il} - \s{ik} - \s{jl} $.  Using this constraint for all the possible combinations of 4 non-coincident endpoints, we obtain $c_{ij} = \pm 1$, depending on the positioning of $i$ and $j$.  
In particular, $c_{ij} = (-1)^{i+j+1}$.
Hence the correct generalisation of the entanglement entropy \req{xanswer}, extended to $n/2$ distinct intervals is 
\begin{equation}
x = \mathop{ \sum_{{i,j=1}}^n}_{j>i} \, (-1)^{i+j+1} \, \s{ij} \ .
\Label{xlinansatzn}
\end{equation}	
This indeed confirms the corresponding expression found by CFT methods \cite{Calabrese:2004eu}.

\section{Disconnected regions in $2+1$ dimensional QFTs}
 \label{disconnectd}

In the previous section we have used bulk techniques to derive the expression for the entanglement entropy of a general disconnected region (composed of $m$ finite intervals) in $1+1$ dimensional QFT.  While the bulk derivation was quite simple, we could have nevertheless used an appropriate uniformization map to reduce the problem to one of computing twist operator correlation functions in a CFT \cite{Calabrese:2004eu} 
and obtained the same result (at least for simple conformally invariant states). 
However, these techniques are applicable only in $1+1$ dimensional CFT, and it is not presently clear what is the analogous expression for entanglement entropy of non-simply connected regions in higher dimensions.  Therefore in this section we explore what the bulk method can tell us about this situation.  For simplicity, we focus on $2+1$ dimensional CFT; so the regions and the bulk minimal surfaces in question are two dimensional.

The holographic dual of entanglement entropy (in some static state of the CFT) of a simply connected region on the boundary is the area of a bulk minimal surface anchored on the boundary of this region.  This relation is true in all dimensions, as are the strong sub-additivity properties of entanglement entropy; indeed, it is easy to see that the proof of \cite{Headrick:2007km} is valid in all dimensions.
Therefore, for situations where the interval structure of \fig{minsurfX} can be straightforwardly generalised to two-dimensional regions, the same reasoning as in \sec{disconnect} above will apply here as well.  However, a simple generalisation from intervals to regions will not hold in general; indeed, such a situation occurs only in a special, limiting, case.

\begin{figure}
\begin{center}
 \includegraphics[width=14cm]{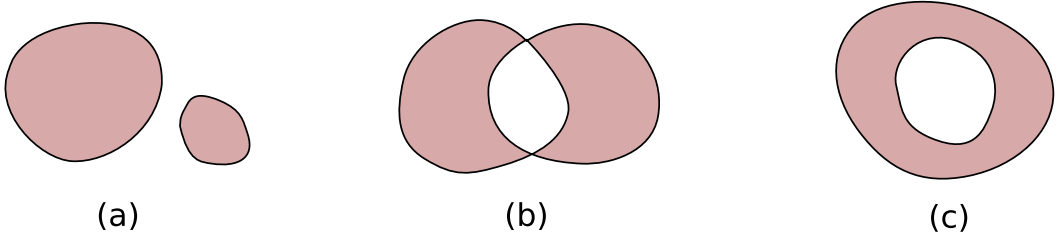}
\caption{Three basic types of finite 2-dimensional non-simple regions $X$ (shaded). Note that case (b) is special, and can be deformed into either of the general cases (a) and (c).} 
\label{twoDregions}
\end{center}
\end{figure}
This is illustrated in \fig{twoDregions}, where we indicate the basic types of non-simply connected regions.  In particular, generic regions whose boundary is not a single non-self-itersecting closed curve are either composed of disjoint simple regions as in case (a), or contain non-contractible curves as in case (c), or some combination thereof.  The intermediate special case which interpolates between (a) and (c) is the case (b) where two simple regions touch at two isolated points.  As we argue below, this is the case for which we can apply our formula for the entanglement entropy \req{xanswer}.   This is because the boundary of region $X$ in (b) naturally defines six minimal surfaces, in direct analogy to \fig{minsurfX}.  On the other hand, both (a) and (c) only define three possible minimal surfaces; two which are homologous to the simple regions and one which in case (a) looks like a handle connecting the shaded regions, and in case (c) is homologous to the shaded annulus.
However, since the configuration (b) can be `resolved' into either (a) or (c), we can use it to conjecture the form of entanglement entropy for the general case as well.\footnote{
Note that all three cases have the same level of complexity, despite case (c) involving only a single connected component, because in each case $X$ is bounded by two curves.
}

Let us focus on the case of the non-simple region $X$ having only two simply connected components, \ie, the two-dimensional analog of the set-up of \sec{disconnecttwo}.
Applying the reasoning of the previous $1+1$ dimensional case requires that the boundary of the non-simple region $X$, generalising the four points $\pt_i$, has four components, say labeled by the four curves $\Cv_i$ (\cf\ \fig{twoDregsCs}b) which enclose six simple regions $\R{ij}$ and to which we can associate the six minimal surfaces $\s{ij}$.  This in turn means that all curves are either infinite or meet at both endpoints.
In particular, for $\s{ij}$ given by the bulk minimal surface anchored on the boundary by the bounding curves $\Cv_i$ and $\Cv_j$ (which form a simple region $\R{ij}$), we obtain as the entanglement entropy of the region $x$ in \fig{twoDregions}b:
\begin{equation}
x = \s{12} + \s{23} + \s{34} +  \s{14}  - \s{13} - \s{24} \ .
\Label{xanswertwo}
\end{equation}	

The simplest type of set-up for which we can calculate the entanglement entropy explicitly is the case of two infinite strips -- \ie, region $X$ of \fig{minsurfX} smeared in one extra direction.  Clearly, we can employ the same reasoning as in \sec{disconnecttwo}, so the entanglement entropy density of $X$ is given by the entanglement entropy densities $\s{ij}$ by the expression \req{xanswertwo}. 
 Note that even though the extra direction is translationally invariant, the individual $\s{ij}$'s are actually different from the one-dimensional case, because of the warp factor associated with the extra direction.  Therefore, the entanglement entropy formula would be extremely difficult to obtain directly in the CFT.

While equation \req{xanswertwo} provides a formula for the entanglement entropy of regions defined by four curves, as in \fig{twoDregions} case (b),  such configurations are highly non-generic in the set of non-simple regions.  The generic cases are given by only two curves, as in  \fig{twoDregions} cases (a) or (c).  The three minimal surfaces defined by these two curves are insufficient to determine the entanglement entropy, as we can see either by taking the limiting case (b) or by noting that the UV divergence for the unregulated entanglement doesn't scale appropriately for any non-trivial superposition of the areas of the minimal surfaces in question.\footnote{
The authors of \cite{Hirata:2006jx} give a very elegant proposal which circumvents the UV problem, while respecting sub-additivity.  In particular, for $X$ bounded by non-intersecting curves $\Cv_{\alpha}$ and $\Cv_{\beta}$, with the minimal surfaces $ s_{\alpha}$, $s_{\beta}$, and  $s_{\alpha\beta}$, their conjecture corresponds to $x = \min \{ s_{\alpha}+s_{\beta} , s_{\alpha\beta}\}$.  However, we believe that this simple prescription cannot hold in general: for example, for very elongated regions in case (a), which limit to the two infinite strips discussed above, we find disagreement from the expected answer: 
$x = \s{12} + \s{23} + \s{34} +  \s{14}  - \s{13} - \s{24} \ne \min \{ \s{12} + \s{34} ,  \s{23} + \s{14} \}$.  (Although there may be special limits or special states for which the two prescriptions are equivalent, one can easily confirm the inequality for the vacuum state and general strips.)
We thank Tadashi Takayanagi for useful discussion on this point.
}  In the following subsection, we elaborate on these cases, and offer a bold conjecture for finding the entanglement entropy. 

\subsection{Conjecture for generic non-simple regions}
 \label{conjecture}

\begin{figure}
\begin{center}
 \includegraphics[width=16cm]{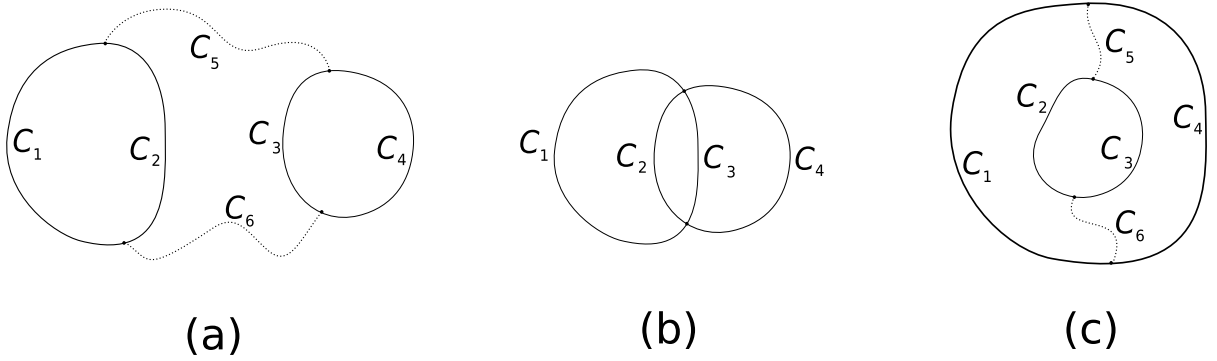}
\caption{Three basic types of finite 2-dimensional non-simple regions $X$ (shaded). Note that case (b) is special, and can be deformed into either of the general cases (a) and (c).} 
\label{twoDregsCs}
\end{center}
\end{figure}

Since we have a prescription for the entanglement entropy for three adjoining regions defining six simple regions and therefore six corresponding minimal surfaces as in \fig{twoDregions} case (b), let us try to express the more general two-curve cases (a) and (c) of \fig{twoDregions},  in a similar manner.
This involves joining the two curves by two more `auxiliary' curves, which simultaneously splits each of the original curves into two components with endpoints at the intersections.  We denote these as indicated in \fig{twoDregsCs}: in case (a), the two disjoint regions are bounded by $\left(\Cv_1 + \Cv_2\right)$ and $\left(\Cv_3 + \Cv_4\)$, while in case (c), the annulus is bounded by $\left(\Cv_1 + \Cv_4\right)$ and $\left(\Cv_2+\Cv_3\right)$.  The two auxiliary curves which join $\{\Cv_1,\Cv_2,\Cv_3,\Cv_4\}$ are denoted by $\Cv_5$ and $\Cv_6$.  The special case (b) then corresponds to vanishing $\Cv_5$ and $\Cv_6$.  In general, we restrict the auxiliary curves $\Cv_5$ and $\Cv_6$ to have no intersections.

Furthermore, to make the notation more explicit, let us denote the simple regions formed by joining the connected curves $\(\Cv_i+\Cv_j+\Cv_k+\Cv_l\)$ by $\R{ijkl}$, and the corresponding entanglement entropy $\EE(\R{ijkl}) = \s{ijkl}$.  So in the first case,  \fig{twoDregsCs}a, we have the six simple regions 
$\R{12}$, $\R{1536}$, $\R{1546}$, $\R{2536}$, $\R{2546}$, and $\R{34}$; and correspondingly, the entanglement entropies 
$\s{12}$, $\s{1536}$, $\s{1546}$, $\s{2536}$, $\s{2546}$, and $\s{34}$.  
Similarly, in the other case,  \fig{twoDregsCs}c, we have
$\R{1526}$, $\R{1536}$, $\R{14}$, $\R{23}$, $\R{2546}$, and $\R{3546}$; and 
correspondingly, the entanglement entropies 
$\s{1526}$, $\s{1536}$, $\s{14}$, $\s{23}$, $\s{2546}$, and $\s{3546}$.
Notice that in each case there is the third minimal surface anchored on the original non-simple regions; $\s{(12)(34)}$ for case (a) and $\s{(14)(23)}$ for case (c).
These are bounded by $\s{1546} + \s{2536}$ in case (a)  and $\s{1526} + \s{3546}$ in case (c).  In the special case (b) these naturally reduce to
$\s{14} + \s{23}$ and $\s{12} + \s{34}$, respectively.

\paragraph{Conjecture:}  Since the entanglement entropies cannot depend on our choice of the auxiliary curves $\Cv_5$ and $\Cv_6$, we {\it minimise} over all such possible configurations, namely:
\begin{eqnarray}
 x &=& \inf_{\Cv_5,\Cv_6} 
 \left\{ \s{12} -\s{1536} + \s{1546} + \s{2536} - \s{2546} + \s{34} \right\}
 \qquad \rm{for \ case \ (a)} \ , \nonumber \\
 x &=& \inf_{\Cv_5,\Cv_6} 
\left\{ \s{1526} -\s{1536} + \s{14} + \s{23} - \s{2546} + \s{3546} \right\}
 \qquad \rm{for \ case \ (c)} \ .
\Label{xanswerac}
\end{eqnarray}	
Note that the partition of the original two curves into the curves $\Cv_i$ for $i = 1, \ldots, 4$ also depends on the choice of  $\Cv_5$ and $\Cv_6$, in particular their endpoints.  Thus our prescription involves choosing two pairs of points on each of the original curves and two curves connecting these points, finding the entanglement entropy using the previously-derived expression \req{xanswertwo}, and minimising over all allowed configurations.

We have chosen the notation in \fig{twoDregsCs} so as to make the limiting case of both (a) and (c) coincide with the special case (b).  The price we pay for this choice is that we have to treat (a) and (c) separately; in particular in  \req{xanswerac} we write distinct expressions for each case.  However, there is a simplifying symmetry that our system respects: in particular, by relabeling $\Cv_2 \tot \Cv_4$ in case (c), the two expressions in \req{xanswerac} become identical.  Thus, we could have written only one formula instead which covers all cases; nevertheless, to help visualising how the various minimal surfaces morph into each other as the case (b) is deformed into either (a) or (c), we will adhere to the notation introduced above.

\paragraph{Support for the conjecture:}
The above conjecture, albeit reasonably natural, has been only minimally motivated. Unfortunately, at present we can furnish neither a derivation, nor an explicit test.  Instead, we offer some supporting evidence for it and leave a detailed analysis for the future.  The main consistency checks are based mostly on the observations that the ultraviolet divergences cancel out correctly, and that the expressions \req{xanswerac} reduce correctly to the known cases in specific limits.
\begin{itemize}
\item
As pointed out previously, we do not expect to obtain an expression for entanglement entropy solely in terms of the three available minimal surfaces; hence we need to furnish more information.  Obtaining the requisite number of minimal surfaces seems the most natural way to proceed.  In particular, in the limiting case (b), the six surfaces $\{\s{12}, \s{13}, \s{14}, \s{23}, \s{24}, \s{34}\}$ comprise the minimal amount of information we need to supply (without specifying the QFT state) in order to determine the entanglement entropy.  On the other hand, the answer can not depend on the extra supplied information.  The expression \req{xanswerac} is therefore the simplest, minimal ansatz.  
\item
Note that the infimum ansatz of  \req{xanswerac} is not a-priori the only possible way to satisfy the above requirements of supplying enough information and the final answer independent of the extra supplied information.  For instance, we could instead consider the supremum over all configurations, or more generally, some weighted sum of all configurations.  However, we now argue that these do not appear correct.  To see that the supremum is incorrect, consider \eg\ case (a) with $\Cv_5$ and $\Cv_6$ coinciding.  Then we would have $x \ge \s{12} + \s{34}$, which is in manifest contradiction with \req{xupcup}, unless $x = \s{12} + \s{34}$.  But the latter leads to the unphysical result that the entanglement entropy is independent of the separation between the regions.\footnote{It is however possible as noted in \cite{Kitaev:2005dm} that suitable linear combinations of entanglement entropy can be made insensitive to the auxiliary curves $\Cv_5$ and $\Cv_6$. } The possibility of expressing $x$ in terms of some weighted sum of the expression $\{ \ldots \}$ in \req{xanswerac} is more difficult to rule out; however we note that this seems to require too much information (namely the weighing factor for each possible configuration).\footnote{In fact the simplest possibility  is that the term in curly braces in \req{xanswerac} be independent of $\Cv_5$ and $\Cv_6$.  However, this is unlikely as can be seen using the following argument by contradiction: If the above were  true, we could in particular take $\Cv_5$ and $\Cv_6$ to coincide and deduce that $x = \s{12}+\s{34}$, independent of the separation between $\R{12}$ and $\R{34}$. To see this cannot happen, let $\Cv_2$ and $\Cv_3$ coincide. Then the region $X$ is simply connected and $x = \s{14}$ for any state (which in general is different from  $\s{12} + \s{34}$).}
\item
A more quantitative way to motivate \req{xanswerac} is to consider the ultraviolet divergences of the unregulated entanglement entropies.  The leading divergence comes from the boundary term, and therefore is proportional to length of the bounding curves.  By construction, we therefore expect that the leading divergent part of $x$ is proportional to the length of $\Cv_1+\Cv_2+\Cv_3+\Cv_4$.  This is indeed exactly reproduced by the expressions in \req{xanswerac}, as can be seen by noting that each subscript $i$ corresponds to $\Cv_i$ and then counting the subscripts.   For example, the first expression in $\{ \ldots \}$ of \req{xanswerac} gives the UV divergence proportional to  $( \Cv_1 + \Cv_2 - \Cv_1 - \Cv_5 -\Cv_3 -\Cv_6 + \Cv_1 + \Cv_5 + \Cv_4 + \Cv_6 + \Cv_2 + \Cv_5 + \Cv_3 + \Cv_6 - \Cv_2 - \Cv_5 - \Cv_4 - \Cv_6 + \Cv_3 + \Cv_4 )
= \Cv_1+\Cv_2+\Cv_3+\Cv_4$.
\item
Furthermore, as remarked above, both expressions in \req{xanswerac} reduce correctly to the limiting case (b).  Although it is manifest that by shrinking $\Cv_5$ and $\Cv_6$ to a point, the expressions $\{ \ldots \}$ in \req{xanswerac} coincide with \req{xanswertwo}, we still need to show that this corresponds to the infimum over all (in this limit closed) curves $\Cv_5$ and $\Cv_6$.  Nevertheless, the latter  follows from the observation that the minimal surfaces bounded by $\Cv_5$ and $\Cv_6$ cancel out in \req{xanswerac}.
\item 
Since $x$ is an infimum over a set of configurations of $\Cv_5$ and $\Cv_6$, by evaluating specific configurations, the expression \req{xanswerac} automatically yields upper bounds for $x$.
For example, by taking coincident $\Cv_5$ and $\Cv_6$ in case (a), we get an upper bound which coincides precisely with the previously obtained upper bound on $x$, namely $x \le \s{12} + \s{34}$.  By the symmetry of interchanging $\Cv_2$ and $\Cv_4$ and case (a) with case (c), we likewise obtain the other upper bound $x \le \s{14} + \s{23}$.  
\item 
Finally, we can easily check that in the limit of $X$ being composed of only one simple region, we obtain the correct $x$.  Specifically, for case (a), consider \eg\ the case of $\Cv_3$ and $\Cv_4$ coinciding.  (A special case of this is when $\Cv_3$ and $\Cv_4$ both shrink to a point; however our argument applies more generally.)  Then $\s{34}=0$, $-\s{1536}+\s{1546}=0$, and $\s{2536}-\s{2546}=0$, so \req{xanswerac} reduces to $x = \inf_{\Cv_5,\Cv_6} \{\s{12}\} = \s{12}$, as required.  Similarly, in case (c), coincident $\Cv_2$ and $\Cv_3$ gives $x = \s{14}$.  
\end{itemize}

Above, we have explained the motivation for the conjecture that the entanglement entropy for a general non-simple regions of \fig{twoDregions} is given by \req{xanswerac}.  However, this implicitly assumes that the entanglement entropy does have a geometric description, in particular that this can be expressed as a function of minimal surface areas of particular regions. It would be interesting to derive this result from first principles, which we hope to return to in the future.

\section{Discussion}
 \label{discussion}

We have presented a simple derivation for the entanglement entropy of disconnected regions in 1+1 dimensional quantum field theories, by exploiting the holographic prescription for computing the same. The result, \req{xanswer}, is consistent with the formulae known for 1+1 CFTs and generalizes them to the non-conformal realm. In deriving this result, we crucially exploited the definition of minimal surfaces and the strong sub-additivity property of entanglement entropy (which also follows from the minimal surface prescription \cite{Headrick:2007km}).  Our derivation consisted of two separate arguments, both yielding the same answer; one employed a natural ansatz and obtained the requisite formula by requiring that it reduce to correct expression in specific limits, while the other relied on self-consistency and used the constraints to obtain coinciding upper and lower bounds on the desired expression.
One intriguing aspect of our derivation are the entropy inequalities \req{sijklmnn}, which appear when we have more than two disconnected regions. These have not hitherto been discussed in the field theory literature and appear to be natural analogs of strong sub-additivity in the general situation.

We then proposed a conjecture for the formula in higher dimensions, concentrating specifically on the 2+1 dimensional case.   When the regions in question touch at two isolated points, our proposal \req{xanswertwo} is a natural generalization of the 1+1 dimensional result. However, as discussed in \sec{disconnectd} one can have more complicated set-up; we conjecture that the entanglement entropy in such circumstances is obtained by adding `virtual regions' to reduce the problem to the previous case, and then minimizing over the space of possibilities for such regions; see \req{xanswerac}. While we have not provided any concrete evidence for our conjecture, we have shown that it is consistent with known data. In particular, we have argued that it correctly captures the leading divergence of the entanglement entropy (the area law)  and reduces to the known results when some components are collapsed to obtain connected regions. It would be very interesting to prove this result from first principles.

We have noted in \sec{disconnectd} that while \cite{Hirata:2006jx} propose a far simpler expression for the entanglement entropy of a region bounded by two curves (namely cases (a) and (c) of \fig{twoDregions}), involving only the three naturally defined minimal surfaces, this doesn't seem to reduce to the requisite result for the infinite strip case.  However, one of the motivations of \cite{Hirata:2006jx} was based on the close analogy with Wilson loop computations. Recall that Wilson loop expectation values in the AdS/CFT context are also given by the regulated area of a bulk minimal surface \cite{Maldacena:1998im,Rey:1998ik} which corresponds to a string world-sheet.  As discussed in  \cite{Gross:1998gk}, the correlation function of two Wilson loops has a transition in the dominant world-sheet from the connected surface ending on the two loops to a disconnected surface ending on each individual loop. However, one important distinction in this situation is the fact that string world-sheeets are oriented (for unitary gauge group). It is therefore plausible that the Wilson loop calculation receives contributions from only two distinct configurations, while the entanglement entropy is sensitive to the correlations across various regions as given by \req{xanswerac}.

So far, in all of our discussion, we have assumed that given a simple region $\R{ij}$ on the boundary, there exists a corresponding bulk minimal surface $\s{ij}$ homologous to $\R{ij}$.  While this is true for the vacuum state (and should be true for any pure state which is sufficiently near the vacuum state, essentially because we expect the minimal surface to exist close to the AdS boundary), $\s{ij}$ is not a-priori guaranteed to exist for a general state. In this context it is useful to recall the examples discussed in \cite{Nishioka:2006gr, Klebanov:2007ws,Faraggi:2007fu}, where the bulk surface contributing to the entanglement entropy changes discontinuously as the length of the boundary region is varied (in a fashion similar to phase transitions involving Wilson loops in thermal AdS/CFT \cite{Witten:1998zw}). However, the local existence of the minimal surface,  coupled with the requirement that we obtain the correct dependence on the UV cut-off for local QFTs,  indicates that the formulae we give are correct, but allow the possibility of the minimal surfaces jumping discontinuously under continuous deformations of the boundary regions.
 
In fact, if we allow for the minimal surfaces to be topologically nontrivial, our conjecture \req{xanswerac} naturally contains the apparently mysterious minimal surfaces anchored on both of the original curves and nowhere else, denoted $\s{(12)(34)}$ and $\s{(14)(23)}$ above.  For example, the former may arise from $\s{1546}$ when $\Cv_5$ and $\Cv_6$ coincide {\it and} this surface has smaller area than the more natural $\s{12}+\s{34}$ configuration.  In other words, as $\Cv_5$ and $\Cv_6$ are brought close enough together, the minimal surface $\s{1546}$ may develop a handle.  It would be interesting to explore the topological restrictions, in particular the requirement of the surfaces being homologous to the corresponding boundary regions, in this context.

An important open problem that we have not discussed in the present work is to generalize the entanglement entropy formulae to the case with time dependence. As discussed in \cite{Hubeny:2007xt}, for general non-static situations one needs to replace the minimal surface by an extremal surface in the bulk.\footnote{This is also necessary for stationary spacetimes, as the timelike Killing field at infinity is not hypersurface orthogonal.} Unfortunately, the arguments given above (or indeed those used in \cite{Headrick:2007km}) do not carry over to the time dependent case, since the extremal surfaces of interest do not lie on a single spacelike surface in the bulk.
This prevents direct comparison of the areas of the surfaces; indeed even the simple inequalities \req{sijk} are harder to obtain in this case owing to the possibility that one can reduce the area of surfaces by wiggling them in the timelike direction. It would be interesting to establish sub-additivity explicitly from the geometrical viewpoint (thereby lending further credence to the proposal of \cite{Hubeny:2007xt}) and derive the formulae for disconnected regions.

\section*{Acknowledgements}
 \label{acks}
It is a pleasure to thank Joan Camps and Tadashi Takayanagi for discussions and comments on the manuscript. This work is supported in part by STFC.

\bibliography{EE_disconnected}

\providecommand{\href}[2]{#2}\begingroup\raggedright\begin{thebibliography}{10}

\bibitem{Srednicki:1993im}
M.~Srednicki, ``Entropy and area,'' {\em Phys. Rev. Lett.} {\bf 71} (1993)
  666--669,
\href{http://www.arXiv.org/abs/hep-th/9303048}{{\tt hep-th/9303048}}.

\bibitem{Kabat:1994vj}
D.~Kabat and M.~J. Strassler, ``A Comment on entropy and area,'' {\em Phys.
  Lett.} {\bf B329} (1994) 46--52,
\href{http://www.arXiv.org/abs/hep-th/9401125}{{\tt hep-th/9401125}}.

\bibitem{Ryu:2006bv}
S.~Ryu and T.~Takayanagi, ``Holographic derivation of entanglement entropy from
  AdS/CFT,'' {\em Phys. Rev. Lett.} {\bf 96} (2006) 181602,
\href{http://www.arXiv.org/abs/hep-th/0603001}{{\tt hep-th/0603001}}.

\bibitem{Ryu:2006ef}
S.~Ryu and T.~Takayanagi, ``Aspects of holographic entanglement entropy,'' {\em
  JHEP} {\bf 08} (2006) 045,
\href{http://www.arXiv.org/abs/hep-th/0605073}{{\tt hep-th/0605073}}.

\bibitem{Hubeny:2007xt}
V.~E. Hubeny, M.~Rangamani, and T.~Takayanagi, ``A covariant holographic
  entanglement entropy proposal,'' {\em JHEP} {\bf 07} (2007) 062,
\href{http://www.arXiv.org/abs/0705.0016}{{\tt 0705.0016}}.

\bibitem{Bousso:1999xy}
R.~Bousso, ``A Covariant Entropy Conjecture,'' {\em JHEP} {\bf 07} (1999) 004,
\href{http://www.arXiv.org/abs/hep-th/9905177}{{\tt hep-th/9905177}}.

\bibitem{Fursaev:2006ih}
D.~V. Fursaev, ``Proof of the holographic formula for entanglement entropy,''
  {\em JHEP} {\bf 09} (2006) 018,
\href{http://www.arXiv.org/abs/hep-th/0606184}{{\tt hep-th/0606184}}.

\bibitem{Emparan:2006ni}
R.~Emparan, ``Black hole entropy as entanglement entropy: A holographic
  derivation,'' {\em JHEP} {\bf 06} (2006) 012,
\href{http://www.arXiv.org/abs/hep-th/0603081}{{\tt hep-th/0603081}}.

\bibitem{Nishioka:2006gr}
T.~Nishioka and T.~Takayanagi, ``AdS bubbles, entropy and closed string
  tachyons,'' {\em JHEP} {\bf 01} (2007) 090,
\href{http://www.arXiv.org/abs/hep-th/0611035}{{\tt hep-th/0611035}}.

\bibitem{Klebanov:2007ws}
I.~R. Klebanov, D.~Kutasov, and A.~Murugan, ``Entanglement as a Probe of
  Confinement,''
\href{http://www.arXiv.org/abs/0709.2140}{{\tt 0709.2140}}.

\bibitem{Faraggi:2007fu}
A.~Faraggi, L.~A. Pando~Zayas, and C.~A. Terrero-Escalante, ``Holographic
  Entanglement Entropy and Phase Transitions at Finite Temperature,''
\href{http://www.arXiv.org/abs/0710.5483}{{\tt 0710.5483}}.

\bibitem{Calabrese:2004eu}
P.~Calabrese and J.~L. Cardy, ``Entanglement entropy and quantum field
  theory,'' {\em J. Stat. Mech.} {\bf 0406} (2004) P002,
\href{http://www.arXiv.org/abs/hep-th/0405152}{{\tt hep-th/0405152}}.

\bibitem{Holzhey:1994we}
C.~Holzhey, F.~Larsen, and F.~Wilczek, ``Geometric and renormalized entropy in
  conformal field theory,'' {\em Nucl. Phys.} {\bf B424} (1994) 443--467,
\href{http://www.arXiv.org/abs/hep-th/9403108}{{\tt hep-th/9403108}}.

\bibitem{Cardy:2007mb}
J.~L. Cardy, O.~A. Castro-Alvaredo, and B.~Doyon, ``Form factors of
  branch-point twist fields in quantum integrable models and entanglement
  entropy,''
\href{http://www.arXiv.org/abs/0706.3384}{{\tt 0706.3384}}.

\bibitem{Headrick:2007km}
M.~Headrick and T.~Takayanagi, ``A holographic proof of the strong
  subadditivity of entanglement entropy,''
\href{http://www.arXiv.org/abs/0704.3719}{{\tt 0704.3719}}.

\bibitem{Lieb:1973cp}
E.~H. Lieb and M.~B. Ruskai, ``Proof of the strong subadditivity of
  quantum-mechanical entropy,'' {\em J. Math. Phys.} {\bf 14} (1973)
1938--1941.

\bibitem{Lieb:1973lr}
E.~H. Lieb and M.~B. Ruskai, ``A fundamental property of quantum-mechanical
  entropy,'' {\em Phys. Rev. Lett.} {\bf 30} (1973) 434 -- 436.

\bibitem{Lieb:1974qr}
E.~H. Lieb, ``Some Convexity and Subadditivity Properties of Entropy,'' {\em
  Bull. Amer. Math. Soc.} {\bf 81} (1975), no.~1, 1--13. Invited talk given at
  A.M.S. Meeting, M.I.T. Oct 27, 1973.

\bibitem{Araki:1970ba}
H.~Araki and E.~H. Lieb, ``Entropy inequalities,'' {\em Commun. Math. Phys.}
  {\bf 18} (1970)
160--170.

\bibitem{Hirata:2006jx}
T.~Hirata and T.~Takayanagi, ``AdS/CFT and strong subadditivity of entanglement
  entropy,'' {\em JHEP} {\bf 02} (2007) 042,
\href{http://www.arXiv.org/abs/hep-th/0608213}{{\tt hep-th/0608213}}.

\bibitem{Fursaev:2007sg}
D.~V. Fursaev, ``Entanglement Entropy in Quantum Gravity and the Plateau
  Problem,''
\href{http://www.arXiv.org/abs/0711.1221}{{\tt 0711.1221}}.

\bibitem{Kitaev:2005dm}
A.~Kitaev and J.~Preskill, ``Topological entanglement entropy,'' {\em Phys.
  Rev. Lett.} {\bf 96} (2006) 110404,
\href{http://www.arXiv.org/abs/hep-th/0510092}{{\tt hep-th/0510092}}.

\bibitem{Maldacena:1998im}
J.~M. Maldacena, ``Wilson loops in large N field theories,'' {\em Phys. Rev.
  Lett.} {\bf 80} (1998) 4859--4862,
\href{http://www.arXiv.org/abs/hep-th/9803002}{{\tt hep-th/9803002}}.

\bibitem{Rey:1998ik}
S.-J. Rey and J.-T. Yee, ``Macroscopic strings as heavy quarks in large N gauge
  theory and anti-de Sitter supergravity,'' {\em Eur. Phys. J.} {\bf C22}
  (2001) 379--394,
\href{http://www.arXiv.org/abs/hep-th/9803001}{{\tt hep-th/9803001}}.

\bibitem{Gross:1998gk}
D.~J. Gross and H.~Ooguri, ``Aspects of large N gauge theory dynamics as seen
  by string theory,'' {\em Phys. Rev.} {\bf D58} (1998) 106002,
\href{http://www.arXiv.org/abs/hep-th/9805129}{{\tt hep-th/9805129}}.

\bibitem{Witten:1998zw}
E.~Witten, ``Anti-de sitter space, thermal phase transition, and confinement in
  gauge theories,'' {\em Adv. Theor. Math. Phys.} {\bf 2} (1998) 505--532,
\href{http://www.arXiv.org/abs/hep-th/9803131}{{\tt hep-th/9803131}}.

\end{thebibliography}\endgroup
\bibliographystyle{utphys}

\end{document}